\documentclass[8.5pt,twoside,twocolumn]{article}
\usepackage[super,sort&compress,comma]{natbib} 
\usepackage[version=3]{mhchem}
\usepackage{balance}
\usepackage{times,mathptm}
\usepackage{sectsty}
\usepackage{graphicx} 
\usepackage{lastpage}
\usepackage[format=plain,justification=raggedright,singlelinecheck=false,font=small,labelfont=bf,labelsep=space]{caption} 
\usepackage{fancyhdr}
\usepackage{color,soul}

\begin{document}

\thispagestyle{plain}
\fancypagestyle{plain}{
\renewcommand{\headrulewidth}{1pt}}
\renewcommand{\thefootnote}{\fnsymbol{footnote}}
\renewcommand\footnoterule{\vspace*{1pt}%
\hrule width 3.4in height 0.4pt \vspace*{5pt}} 
\setcounter{secnumdepth}{5}

\makeatletter 
\renewcommand\@biblabel[1]{#1}            
\renewcommand\@makefntext[1]%
{\noindent\makebox[0pt][r]{\@thefnmark\,}#1}
\makeatother 
\renewcommand{\figurename}{\small{Fig.}~}
\sectionfont{\large}
\subsectionfont{\normalsize} 

\fancyfoot{}
\fancyfoot[RO]{\footnotesize{\sffamily{1--\pageref{LastPage} ~\textbar  \hspace{2pt}\thepage}}}
\fancyfoot[LE]{\footnotesize{\sffamily{\thepage~\textbar\hspace{3.45cm} 1--\pageref{LastPage}}}}
\fancyhead{}
\renewcommand{\headrulewidth}{1pt} 
\renewcommand{\footrulewidth}{1pt}
\setlength{\arrayrulewidth}{1pt}
\setlength{\columnsep}{6.5mm}
\setlength\bibsep{1pt}

\twocolumn[
  \begin{@twocolumnfalse}
\noindent\LARGE{\textbf{Atomistic Simulations of Highly Conductive Molecular Transport Junctions Under Realistic Conditions$^\dag$}}
\vspace{0.6cm}

\noindent\large{\textbf{William R. French,\textit{$^{a}$} Christopher R. Iacovella,\textit{$^{a}$} Ivan Rungger,\textit{$^{b}$} Amaury Melo Souza,\textit{$^{b}$} Stefano Sanvito,\textit{$^{b}$} and Peter T. Cummings$^{\ast}$\textit{$^{a,c}$}}}\vspace{0.5cm}


 \end{@twocolumnfalse} \vspace{0.6cm}

  ]
\noindent\small{\textbf{We report state-of-the-art atomistic simulations combined with high-fidelity conductance calculations to probe structure-conductance relationships in Au-benzenedithiolate(BDT)-Au junctions under elongation. Our results demonstrate that large increases in conductance are associated with the formation of monatomic chains (MACs) of Au atoms directly connected to BDT. An analysis of the electronic structure of the simulated junctions reveals that enhancement in the $s$-like states in Au MACs causes the increases in conductance. Other structures also result in increased conductance but are too short-lived to be detected in experiment, while MACs remain stable for long simulation times. Examinations of thermally evolved junctions with and without MACs show negligible overlap between conductance histograms, indicating that the increase in conductance is related to this unique structural change and not thermal fluctuation. These results, which provide an excellent explanation for a recently observed anomalous experimental result [Bruot \textit{et al., Nature Nanotech.}, 2012, \textbf{7}, 35-40], should aid in the development of mechanically responsive molecular electronic devices.}}
\section*{}
\vspace{-1cm}


\footnotetext{\textit{$^{a}$~Department of Chemical and Biomolecular Engineering, Nashville, TN, USA. E-mail: peter.cummings@vanderbilt.edu}}
\footnotetext{\textit{$^{b}$~School of Physics and CRANN, Trinity College, Dublin 2, Ireland }}
\footnotetext{\textit{$^{c}$~Center for Nanophase Materials Sciences, Oak Ridge National Laboratory, Oak Ridge, TN, USA. }}



While benzene-1,4-dithiolate (BDT) has been widely studied over the years for applications in molecular electronics, \cite{Reed:1997,Xiao:2004,Tsutsui:2009-nanoscale,Song:2009,Bruot:2012,Kim:2011,Sergueev:2010,Pontes:2011,Strange:2010,Romaner:2006,Ke:2005,French:2012,Pu:2010,French:2013} recent discoveries \cite{Song:2009,Bruot:2012,Kim:2011} of its tunable conductance properties have generated a renewed interest in the molecule. These discoveries include counterintutive conductance increases (exceeding an order of magnitude) during elongation of a Au-BDT-Au junction,\cite{Bruot:2012} and a wide conductance window spanning three orders of magnitude.\cite{Kim:2011} The tunability of BDT's conductance is enabled by the proximity of its highest occupied molecular orbital (HOMO) to the Au electrode Fermi level ($\epsilon_{F}$). Relatively small increases in the HOMO level result in resonant (or near resonant) tunneling, which, as demonstrated by Bruot $et$ $al.$ \cite{Bruot:2012} can be achieved through mechanical elongation of a Au-BDT-Au junction. However, elongation does not guarantee increases in conductance, as evidenced by numerous previous Au-BDT-Au studies \cite{Reed:1997,Xiao:2004,Kim:2011} where conductance increases were not reported. The reason for this discrepancy is that the exact location of the HOMO level depends on the structural evolution of a junction, which may be influenced by the experimental setup and/or conditions. In order to make use of the desirable properties of BDT, and to facilitate experimental reproducibility, it is essential to determine the structure(s) responsible for increasing conductance.

Prior computational work \cite{Sergueev:2010,Pontes:2011,Strange:2010,Romaner:2006,Ke:2005} has thus far provided considerable insight into the behavior of BDT and the underlying mechanisms that control conductance. However, most prior computational studies of molecular junctions have adopted simplifications which may ultimately make it difficult to connect the predicted behavior with experiment. For instance, in typical computational studies of BDT, the molecule is  sandwiched between ideal, planar surfaces with an arbitrary initial geometry, and then stretched via geometry optimizations.\cite{Sergueev:2010,Pontes:2011,Romaner:2006,Ke:2005} This approach may not adequately capture many important aspects found in experiment, such as temperature effects, elongation rate effects, and non-ideal tip geometry. In order to better connect theory to experiment and facilitate a deeper understanding of the structure-conductance relationship, a different computational approach is needed that can capture these environmental effects and the stochastic nature of junction formation.  Here, we combine detailed atomistic simulations with first-principles calculations to investigate the structural basis for the conductance behavior of elongating Au-BDT-Au junctions. We perform a statistically significant number of simulations ($>$100) that, unlike previous comptuational studies,\cite{Sergueev:2010,Pontes:2011,Strange:2010,Romaner:2006} accounts for the spontaneous connection of a BDT molecule between two ruptured nanowire tips and thus are significantly less biased by their starting configuration and more representative of experiment.\cite{Bruot:2012} 

\begin{figure}[h]
\centering
  \includegraphics[width=1.8in]{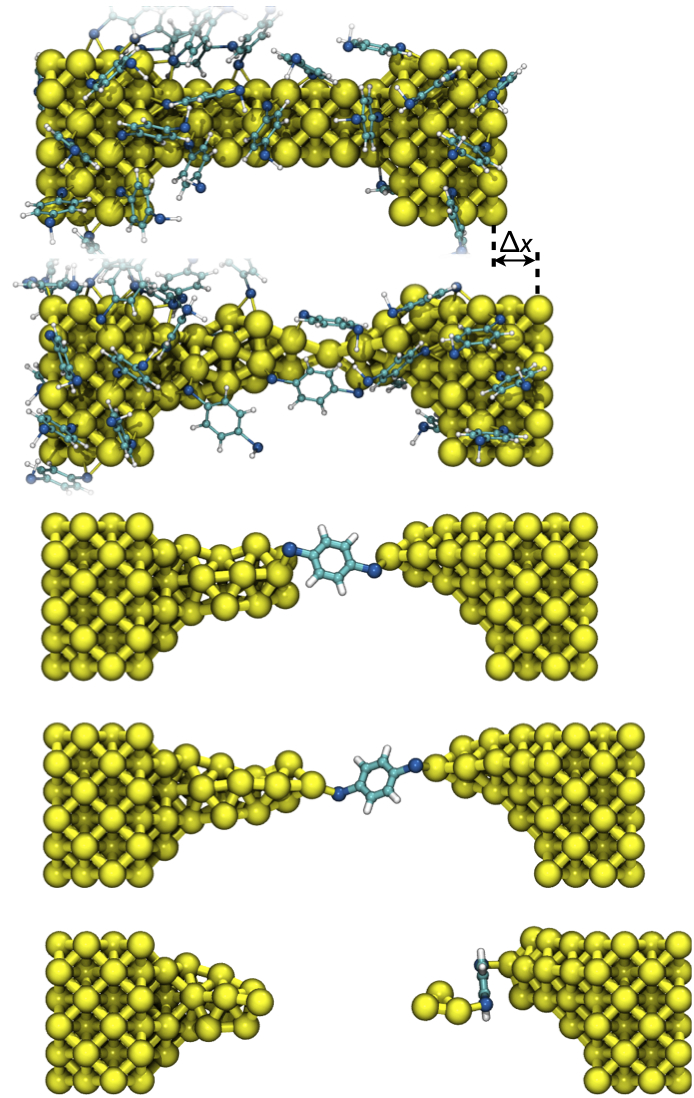}
  \caption{Simulation snapshots of the elongation of a BDT-coated Au nanowire, leading to the formation, elongation, and eventual rupture of a Au-BDT-Au junction. From top to bottom, $\Delta$$x$ = 0.0, 4.0, 8.0, 12.0, and 16.0 \AA. Monolayer molecules are removed after junction formation to isolate electrode geometry effects. All images in this work were rendered using Visual Molecular Dynamics.\cite{Humphrey:1996} \label{fig:methodology}}
\end{figure}

Specifically, in this work, we apply a hybrid molecular dynamics-Monte Carlo (MD-MC) protocol \cite{French:2012,Pu:2010,French:2013} within LAMMPS \cite{Plimpton:1995} and an in-house code to incorporate more realistic environmental features into computational studies of molecular junctions. Au-Au interactions are described using a semi-empirical many-body potential \cite{Cleri:1993} and the S-Au interactions are modeled with multi-site bonding curves parameterized from density functional theory (DFT).\cite{Leng:2007} The simulation procedure is illustrated in Fig. 1 and involves trapping a single BDT molecule between ruptured Au nanowire tips by mechanical deforming a BDT coated nanowire; once a BDT connects between two tips, the junction is evolved via mechanical elongation at finite temperature (77 K), and the atomic coordinates are periodically extracted for use as input to electron transport calculations. This simulation procedure closely models the widely used mechanically controllable break junction experimental method \cite{Reed:1997,Kim:2011,Bruot:2012} and thus should produce experimentally representative configurations. Additionally, this procedure enables us to more efficiently collect a statistically relevant number of configurations than typically used in quantum mechanical simulations, allowing us to capture the ensemble behavior; specifically in this work, we perform 104 independent simulations to model break junction experiments, where we note that recent  quantum mechanical studies \cite{Sergueev:2010,Pontes:2011,Strange:2010,Romaner:2006} considered only at most four structurally distinct junctions. We believe that the geometries predicted by our simulations should better represent break junction experiments than previous computational work \cite{Sergueev:2010,Pontes:2011,Strange:2010,Romaner:2006} due to the inclusion of (1) realistic electrodes, (2) finite temperature effects, (3) better statistics, and (4) junctions that are less biased by starting configuration. For more details about the simulation method and applied force fields please refer to the ESI.\dag

In our break junction simulations, a BDT molecule attaches between two ruptured nanowire tips in 31 out of 104 (30\%) independent runs, in excellent agreement with values reported in Au/BDT break junction experiments (30-40\%).\cite{Xiao:2004} Akin to the experimental situation, the formation of a molecular junction is dictated by the proximity of adsorbed BDT molecules to the nanowire fracture location, and the availability of bonding sites on the Au tip(s). The geometry of a Au tip depends on the structural pathway of the nanowire during the elongation process, and hence can vary substantially between runs.\cite{Coura:2004, Pu:2008, Tavazza:2010, French:2011,Iacovella:2011} Thus, a large number of junction geometries are possible, which we expect to better represent experiment than prior computational studies.

To elucidate the structure-conductance relationship, we calculate the zero-bias conductance of our simulated junctions within the SIESTA \cite{Ordejon:1996} and SMEAGOL \cite{Rocha:2006,Rungger:2008} packages, employing an approximate self-interaction correction (ASIC) \cite{Pemmaraju:2007} method to the local density approximation (LDA), which yields conductance values that more closely match the experimentally most-probable value \cite{Xiao:2004,Tsutsui:2009} than standard DFT \cite{Toher:2008} (although they are still higher by a factor of approximately six). The conductance values we report here for our simulated junctions are generally higher than the most-probable value reported in some experiments,\cite{Xiao:2004,Tsutsui:2009} while they are within the range of values reported in more recent experiments. \cite{Bruot:2012,Kim:2011} The source of the quantitative disagreement between the experimenally measured and theoretically calculated values of Au-BDT-Au conductance is still a debated topic.\cite{DiVentra:2000,Muller:2006,Nichols:2010} As such, here we focus on relative changes in conductance rather than the exact values. A detailed description of the conductance calculations, and results from benchmark calculations validating the conductance calculations and force fields are provided in the ESI.\dag 

\begin{figure*}
  \centering
  \includegraphics[width=6.6in]{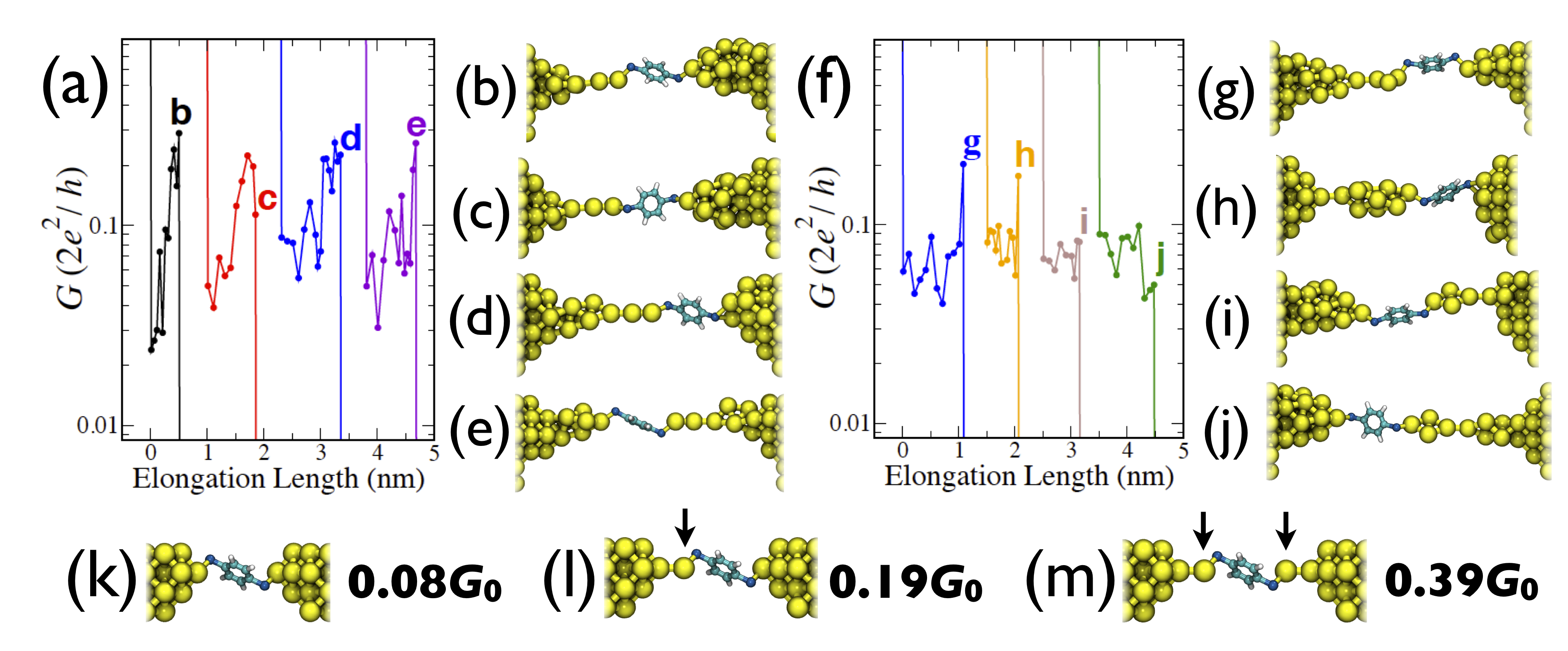}
  \caption{Conductance behavior of Au-BDT-Au junctions undergoing elongation. (a) Four gradually increasing conductance traces, with the corresponding geometries immediately prior to junction rupture shown to the right. The individual traces are offset along the $x$-axis for clarity. At a distance of 0.1 \AA\ prior to junction formation the conductance is assumed to be 1$G_{0}$ while at rupture the conductance is set to 0$G_{0}$. (f) Four relatively flat conductance traces, with the corresponding break geometries shown to the right. (k-m) The effect on conductance of manually connecting Au MACs (indicated with arrows) to BDT.  The BDT is initially connected between two ideal Au(100) tips. A comparison of the transmission and projected density of states is included in the ESI, see Figures S4 and S5.\dag}
  \label{fig:traces}
\end{figure*}

In Fig. 2 we plot the conductance evolution of eight representative Au-BDT-Au junctions under elongation. Distinct differences in the shape of the conductance traces are apparent in Fig. 2; for clarity of presentation, we group similar traces into two separate plots shown in Fig. 2a and Fig. 2f. The curves in Fig. 2a undergo large gradual increases, while the curves in Fig. 2f are relatively flat; we note that both behaviors closely match recent experimental results \cite{Bruot:2012} and that previous idealized computational studies have failed to capture this range of behaviors.\cite{Sergueev:2010,Pontes:2011} Furthermore, these results demonstrate that increases in conductance depend on structure and are not a natural consequence of the increasing potential energy of a junction (see Figure S5 in the ESI\dag).

Interestingly, the breaking geometries shown in Fig. 2b-e all contain a monatomic chain (MAC) of Au atoms directly connected to BDT. Of the 31 simulations resulting in molecular junction formation, 13\% go on to form direct MAC-BDT connections during elongation.  Here we find that each MAC-containing geometry in Fig. 2b-e results in conductance $>$ 0.2$G_{0}$. Closer examination of the entire stretching trajectories reveals that the initial jumps to conductance values larger than 0.1$G_{0}$ coincide with the appearance of a MAC at a BDT-Au interface, suggesting that the BDT-MAC connection is responsible for the increased conductance. To test this, we manually insert MACs at the BDT-Au interface of an ideal junction, as shown in Fig. 2k-m, finding that conductance increases as MACs are inserted into otherwise static geometries. Analysis of the electronic structure of the simulated junctions reveals that MACs broadly (+/- 1 eV) enhance the projected density of states (PDOS; see Figure S7 in the ESI\dag) around $\epsilon_{F}$ for the MAC Au $s$ and $p_{z}$ states, thus inducing a stronger coupling of the molecular states around $\epsilon_{F}$ with the Au electrodes, resulting in a higher transmission. Importantly, this enhancement only occurs when BDT is bonded to a Au atom that bonds with one other Au atom; adatoms or atomically sharp tips (Fig. 2k) do not result in this enhancement (see Figure S7 in the ESI\dag). Previous studies employing ideal, planar electrodes have reported increased conductance for BDT connected to a Au adatom at both electrodes.\cite{Sergueev:2010,Ke:2005} Here, we propose a distinct structural mechanism in which BDT is connected to a MAC at only one of the electrodes. This is supported by atomistic simulations that closely mimic break-junction experiments, providing evidence for the formation of MACs under realistic conditions. Meanwhile, prior studies employed manually generated and/or geometry-optimized (effectively 0 K) structures, providing no evidence of the formation of such structures under dynamic, non-ideal conditions.

In contrast to MACs, other low-coordination electrode structures such as Au-Au$_2$-Au units (see Fig. 2i,j) do not increase conductance (note that Au-Au$_2$-Au  structures have been previously observed in DFT-based studies of Au tips\cite{Li:2007,Tavazza:2010}). Meanwhile, the structures in Fig. 2g,h result in high conductance, but their lifetimes are extremely short; in non-stretching MD simulations the structures shown in Fig. 2g,h remain stable for less than 1.0 ns, which is a time scale that is too short to be measured in experiment. On the other hand, three of the four MACs in Fig. 2b-e remain stable for the complete duration of a 1.0-$\mu$s simulation without stretching (the actual simulated junctions were taken 0.2-0.5 \AA\ of elongation prior to the rupture structures shown in Fig. 2b-e), indicating that MACs may possess sufficient stability to be detected in experiments performed at 77 K.

 \begin{figure}[h!]
\centering
  \includegraphics[width=3.2in]{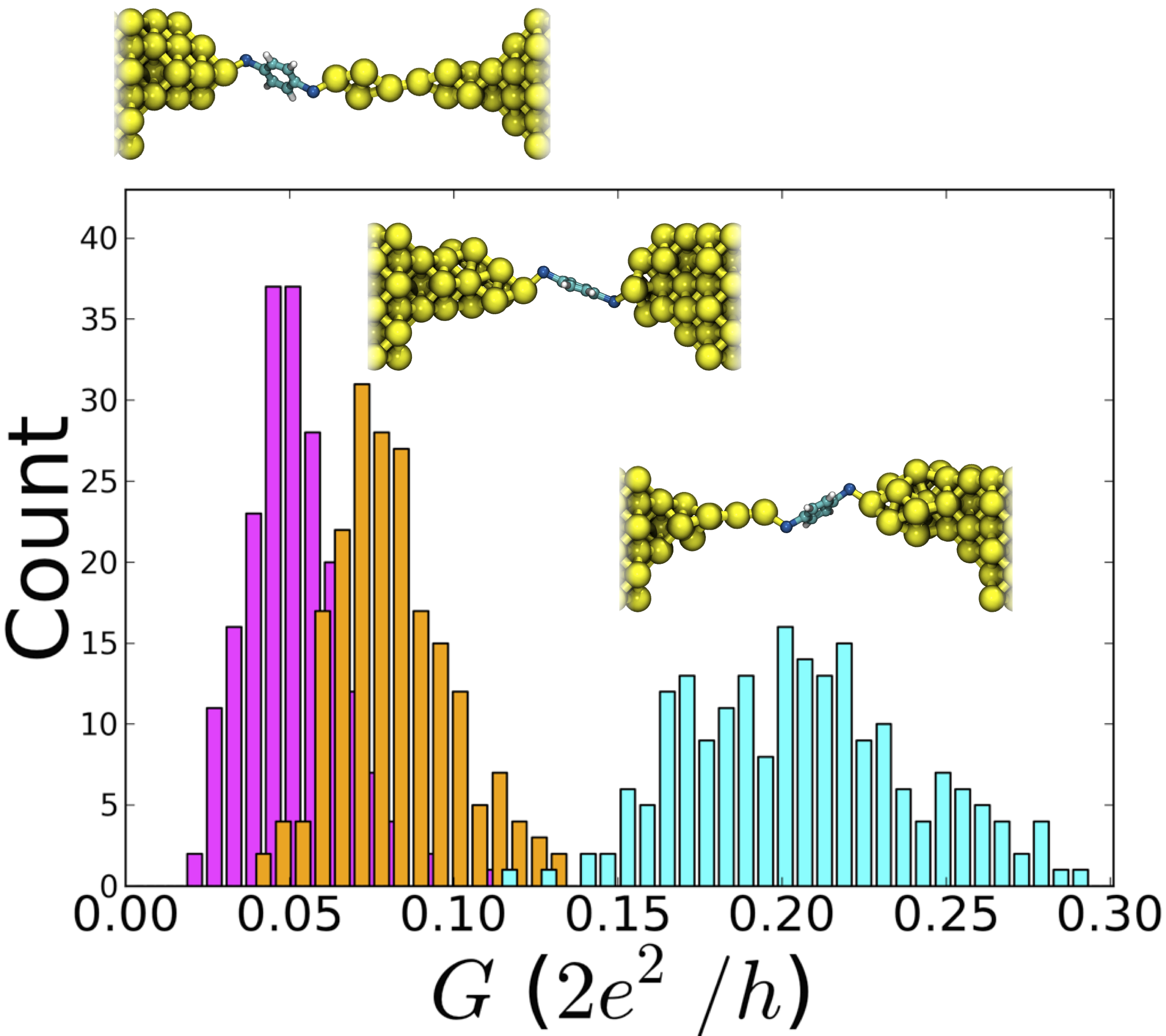}
  \caption{Conductance histograms of three thermally evolving Au-BDT-Au junctions. The right-most junction/histogram corresponds to a system with a MAC, where we see negligible overlap with the non-MAC-containing junctions. The bin width is 0.006$G_{0}$. Standard deviations of the histograms (from left to right) are 0.014$G_{0}$, 0.018$G_{0}$, and 0.034$G_{0}$. \label{fig:histograms}}
\end{figure}

Note that MACs have been observed in experiments \cite{Coura:2004, Yanson:1998, Ohnishi:1998} and in simulations \cite{Pu:2008,Coura:2004,Tavazza:2010} of elongating Au nanowires without BDT. MACs have also been observed in simulations of thiolate-terminated molecules being pulled away from step edges on Au surfaces \cite{Kruger:2002} and in simulations of Au-alkanedithiolate-Au junctions.\cite{Paulsson:2009} Thus, the appearance of MACs in Au/BDT break junction experiments is not without precedent and their formation is not an artifact of our simulation model. However, the significant impact of MACs on molecular conductance has not been previously demonstrated.

To confirm that the conductance increases are not the result of the limited time scale of our simulations, we next compare the conductance distributions of three structurally distinct Au-BDT-Au junctions (see Fig. 3) evolved at 77 K without stretching the junction. Each junction is evolved for 40 ns, with conductance measurements taken every 0.2 ns; this time frame allows for sufficient sampling of the most probable structures and results in converged histograms. The conductance of BDT connected to a MAC is statistically distinct from the two other deformed junctions with almost no overlap between the histograms, indicating that the increases seen in Fig. 2a are indeed associated with their own unique structural motif and not simply a short-lived or improbable configuration. We also observe that the conductance fluctuations for BDT connected to a Au MAC is approximately double that of the other structures due to increased electrode motion. \cite{French:2013} Increased fluctuations in BDT-MAC geometries have been reported previously,\cite{Basch:2005} albeit for unrealistic, manually adjusted geometries, making it difficult to assess their connection to experiment. We also note that the histograms in Fig. 3 are similar to those in experiments by Tsutsui \textit{et al.}, \cite{Tsutsui:2006} where a high-conductance state at 0.1$G_{0}$ exhibited a larger peak width than the peak at lower values (0.01$G_{0}$). 

The similarities between our simulations and the experiments of Bruot $et$ $al.$ \cite{Bruot:2012} suggest that MAC formation is the likely cause of the large (factor of five or more) gradual increases in conductance observed in these experiments. The lack of large conductance increases in experiments performed at 300 K \cite{Bruot:2012} suggests that the structural motif responsible for increased conductance has a distinct thermal dependence ($i.e.$, the structure is unstable at higher temperature). It has been previously \cite{Pu:2008} established that the formation of MACs in mechanically deformed Au nanowires will depend on temperature, where MACs form with the highest frequency and stability at low temperature. We confirm this by performing 298-K simulations without stretching of the MAC geometries in Fig. 2b-e (the actual simulated junctions were taken 0.2-0.5 \AA\ of elongation prior to the rupture structures shown in Fig. 2b-e). In all four cases the junction undergoes thermoactivated spontaneous breakdown in less than 1.0 ns, indicating that MACs are not stable at high temperature. Thus, the thermal instability of MACs explains why large conductance increases were not observed in 300-K experiments.\cite{Bruot:2012} While in the work of Bruot $et$ $al.$ \cite{Bruot:2012} there were no large conductance increases at 300 K, large conductance values were reported in the 300-K conductance histogram, which may have been caused by other significant conformational changes to the junction at 300 K.   

Other factors, such as Au adatoms and strained S-Au bonds, have been shown \cite{Pontes:2011,Sergueev:2010,Romaner:2006,Hoft:2006,Ke:2005} to increase conductance in idealized junctions, and may also play a role under certain conditions. For instance, adatom formation may be important in STM-based break junction experiments \cite{Xiao:2004} where a planar surface serves as one of the electrodes. Meanwhile, the small ($\sim$0.001$G_{0}$) reversible  changes in conductance (with respect to junction compression and elongation) in the work of Bruot $et$ $al.$ \cite{Bruot:2012} are likely caused by reversible structural change such as a strained S-Au bond. On-hollow bonding geometry, \cite{Tsutsui:2006,Kim:2011} high tilt angles, \cite{Haiss:2008,Kim:2011} and S atoms embedded in Au contacts \cite{Arroyo:2011} have all been proposed to explain large values of conductance, but these explanations do not seem likely given the very large, gradual conductance increases observed in the recent work of Bruot $et$ $al.$ \cite{Bruot:2012} or in our computational studies.  Additionally, we note that since the elongated junctions are under tension, it makes it more probable for BDT to adopt an upright geometry with low tilt angle, bonded with a single Au atom at each electrode.   

\begin{figure}[t!]
\centering
  \includegraphics[width=3.5in]{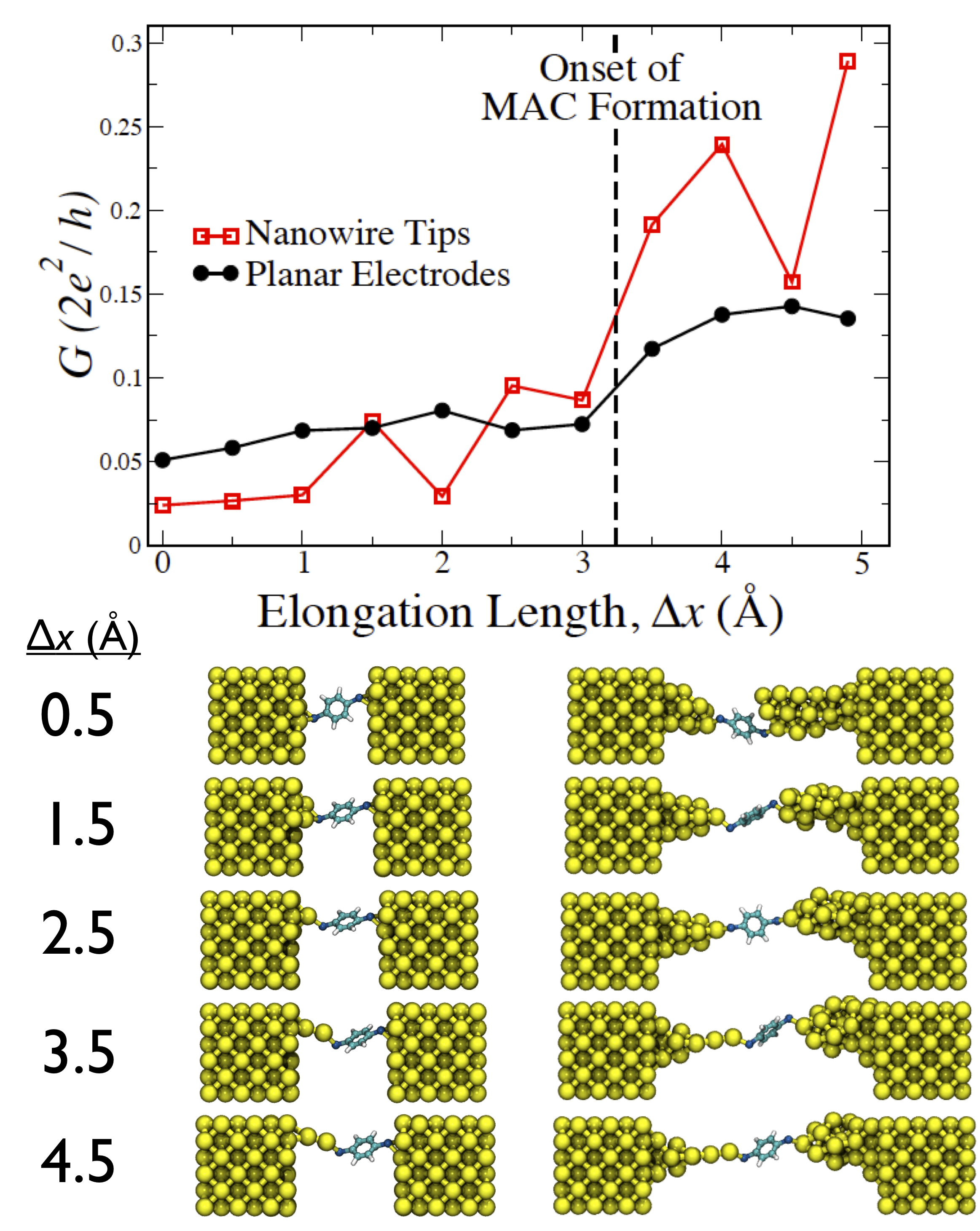}
  \caption{Comparison of the conductance (top) and structural (bottom) evolution of Au-BDT-Au junctions where the electrodes are modeled as (left) nanowire tips and (right) planar surfaces. \label{fig:electrode-comparison}}
\end{figure}

It is also important to note that the formation of MACs in prior theoretical work focusing on flat plates  \cite{Sergueev:2010,Pontes:2011,Strange:2010,Romaner:2006} may be suppressed due to the electrode geometry choice ($i.e.$, flat surfaces).  Flat surfaces could discriminate against the formation of MACs since the \textit{dynamic structural fluxionality} ($i.e.$, lengthening and weakening of Au-Au bonds) of flat Au surfaces is lower than that of nanostructured Au surfaces.\cite{Rashkeev:2007} The use of DFT-based geometry optimizations,\cite{Sergueev:2010,Pontes:2011} where elongation rate and temperature effects are neglected, may also explain the absence of MACs in prior work, since the appearance of MACs has been linked to both the elongation rate and temperature.\cite{Pu:2008} To explore this we perform a Au-BDT-Au elongation simulation using ideal, planar Au(100) surfaces to model the electrodes (see Fig. 4). In this simulation a Au adatom is first pulled from the surface of one of the electrodes, which then evolves into a 2-atom MAC before the junction ruptures. This result supports the important role of temperature in MAC formation, as simulations employing geometry optimizations \cite{Sergueev:2010,Pontes:2011,Strange:2010,Romaner:2006} have not led to MAC formation. In support of our proposed structural mechanism, adatom and MAC formation in our simulations is limited to a single electrode, while prior work \cite{Ke:2005} reported significant conductance increases only when BDT connected to an adatom at \textit{both} electrodes. Although the flat-surface and nanowire junctions exhibit roughly the same ductiliy ($i.e.$, elongation prior to rupture), there are significant differences between the two curves in Fig. 4. Both tend to increase during elongation, however the increase is much larger for the nanowire tips. For the planar electrodes the conductance increases by a factor of almost 3, while for the nanowire tips the conductance increases by over an order of magnitude. The changes in conductance are also more gradual for the planar electrodes, whereas the nanotip data exhibit sharp fluctuations owing to the increased electrode motion.\cite{French:2013}

In summary, by combining realistic simulations of molecular junction formation and elongation with high-fidelity conductance calculations, we have provided important new insight into the conductance behavior of Au-BDT-Au junctions. Namely, we showed that BDT connected directly to a MAC results in enhanced conductance, and is caused by enhancements in the $s$ and $p_{z}$ density of states around $\epsilon_{F}$ in Au MAC atoms. This result offers an excellent explanation for the large, anomalous conductance increases observed in Au-BDT-Au break junction experiments, \cite{Bruot:2012} and may additionally explain the large transmission observed in experiments by Kim and co-workers. \cite{Kim:2011} To further support our conclusion, we show that BDT-MAC structures are stable for long simulation times performed at 77 K, and exhibit very little overlap in their conductance distributions with those of other realistic junction geometries, and thus should be detectable in experiment. These results may aid experimental reproducibility by prompting the development of new strategies for tailoring the structures and properties of molecular junctions, in particular mechanically reponsive devices.\cite{Bruot:2012} Our findings also stress the importance of incorporating realistic environmental conditions into simulation models of molecular junctions for better resolving experimental uncertainty.

\balance

\noindent\large{\textbf{Acknowledgements}}

\vspace{0.3cm}

\noindent\small{WRF acknowledges partial support from the U.S. Department of Education for a Graduate Assistance in Areas of National Need (GAANN) Fellowship under grant number P200A090323; WRF, CRI and PTC acknowledge partial support from the National Science Foundation through grant CBET-1028374. IR, AMS, and SS thank the King Abdullah University of Science and Technology (ACRAB project) for financial support. This research used resources of the National Energy Research Scientific Computing Center (NERSC), which is supported by the Office of Science of the U.S. Department of Energy under Contract No. DE-AC02-05CH11231; specifically, the conductance calculations were performed on NERSC's Carver.}

\footnotesize{
\bibliography{nature-library} 
\bibliographystyle{rsc} }

\end{document}